\documentclass{pazhb}

\usepackage{graphicx} 
\usepackage{amsmath} 
\usepackage{epsfig,graphics} 
\usepackage{rotating} 
\usepackage{ulem}

\usepackage{latexsym}
\usepackage{amsfonts}
\usepackage{soul}
\usepackage{array}
\usepackage{amssymb}
\usepackage{extarrows}

\usepackage{subfig}
\usepackage{wrapfig}
\usepackage{wasysym}
\usepackage{enumitem}
\usepackage{adjustbox}
\usepackage{ragged2e}
\usepackage[svgnames,table]{xcolor}
\usepackage{tikz}
\usepackage{changepage}
\usepackage{setspace}
\usepackage{hhline}
\usepackage{multicol}
\usepackage{tabto}
\usepackage{float}
\usepackage{multirow}
\usepackage{makecell}
\usepackage{fancyhdr}
\usepackage[toc,page]{appendix}
\usepackage[hidelinks]{hyperref}

\usepackage{epstopdf}
\usepackage{pdfpages}

\usepackage{longtable}
\usepackage{dcolumn}
\usepackage{pdflscape}

\usepackage{hyperref}

\usepackage{comment}
\usepackage{ulem}
\usepackage{color}
\usepackage{tcolorbox}
\usepackage{xspace}

\newcommand {\be}{\begin{equation}} 
\newcommand {\ee}{\end{equation}}

\newcommand{\cjaa}{Chin. J. Astron. Astrophys.}

\def\FX0.52{F_{{\rm X},0.5-2}}
\def\LX210{L_{{\rm X},2-10}}

\def\kev2l{L_{2keV}}
\def\l5100A{L_{2500\mathring{A}}}

\begin{document} 

\journalinfo{2020}{46}{11}{726}[733]
\title{Measurement of the Supermassive Black Hole Masses in Two Active Galactic Nuclei by the Photometric Reverberation Mapping Method}
\author{\bf 
  E.A.~Malygin\email{male@sao.ru}\address{1},  E.S.~Shablovinskaya\address{1}, R.I.~Uklein\address{1},  A.A.~Grokhovskaya\address{1}
  \addresstext{1}{Special Astrophysical Observatory Russian Academy of Sciences (SAO RAS), Nizhnii Arkhyz, Karachai-Cherkessian Republic, 369167, Russia}}
\shortauthor{Malygin et al.}  
\shorttitle{Measurement of the supermassive black hole masses} 
\submitted{September 1, 2020; revised October 7, 2020; accepted October 27, 2020}
\begin{abstract}  

We present the results of long-term photometric monitoring of two active galactic nuclei,  2MASX\,J08535955+7700543 (\textit{z} $\sim$ 0.106)  and VII\,Zw\,244 (\textit{z} $\sim$ 0.131), being investigated by the reverberation mapping method in medium-band filters. To estimate the size of the broad line region, we have analyzed the light curves with the JAVELIN code. The emission line widths have been measured using the spectroscopic data obtained at the 6-m BTA telescope of SAO RAS. We give our estimates of the supermassive black hole masses $\lg (M/M_{\odot})$, $7.398_{-0.171}^{+0.153}$, and $7.049_{-0.075}^{+0.068}$, respectively.

\textbf{DOI:} 10.1134/S1063773720110055

\end{abstract}

\section{INTRODUCTION}

One of the important characteristics of a galaxy is the mass of the supermassive black hole (SMBH) located at its center. The SMBH mass is now known to correlate both with the luminosity of the host galaxy and with its stellar velocity dispersion \citep{Gebhardt00,Ferrarese00}. The evolutionary connection between SMBHs and spheroidal stellar components (bulges) of galaxies and with the galactic dark halos is also studied ~\citep[see,~e.g.,][and references therein]{Zasov_TM}.
The SMBH masses in the nuclei of nearby galaxies are measured by studying the dynamics of stars, gas, and maser sources in the gravitational field of the black hole. However, in active galactic nuclei (AGNs) the circumnuclear region is spatially unresolvable and is illuminated by the central source, making a direct study of the dynamics of gravitating matter impossible.

\begin{table*}[!ht]
	\centering
	\caption{Characteristics of the AGNs being investigated. From left to right, the table gives: the object name, the epoch J2000 coordinates, the $V$ magnitude, the redshift $z$, the emission line in which reverberation mapping is performed, and the filters from the SED set used for this purpose, where the number corresponds to the central filter transmission wavelength}
	\begin{tabular}{p{1.5in}p{2.5in}p{0.29in}p{0.38in}p{0.48in}p{0.7in}}
		\hline
		\hline
		\multicolumn{1}{p{1.27in}}{ Object} & 
		\multicolumn{1}{p{1.87in}}{\centering Coordinates \par \centering (RA, Dec, J2000)} & 
		\multicolumn{1}{p{0.29in}}{\centering $V$} & 
		\multicolumn{1}{p{0.38in}}{\centering $z$} & 
		\multicolumn{1}{p{0.48in}}{\centering {\fontsize{11pt}{13.2pt}\selectfont Line}} & 
		\multicolumn{1}{p{0.7in}}{\centering {\fontsize{11pt}{13.2pt}\selectfont Filters} \par \centering {\fontsize{10pt}{12.0pt}\selectfont (line + cont)}} \\
		\hline 
		\multicolumn{1}{p{1.27in}}{ {\fontsize{11pt}{13.2pt}\selectfont 2MASX J08535955+7700543}} & 
		\multicolumn{1}{p{1.87in}}{\centering {\fontsize{11pt}{13.2pt}\selectfont 08\textsuperscript{h}53\textsuperscript{m}59\textsuperscript{s}.4+77$ ^{\circ} $00\arcmin 55\arcsec}} & 
		\multicolumn{1}{p{0.29in}}{\centering 17.0} & 
		\multicolumn{1}{p{0.38in}}{\centering 0.106} & 
		\multicolumn{1}{p{0.48in}}{\centering H$ \alpha $ } & 
		\multicolumn{1}{p{0.7in}}{\centering SED725 \par \centering SED700} \\
		\hline
		\multicolumn{1}{p{1.27in}}{ {\fontsize{11pt}{13.2pt}\selectfont VII Zw 244}} & 
		\multicolumn{1}{p{1.87in}}{\centering {\fontsize{11pt}{13.2pt}\selectfont 08\textsuperscript{h}44\textsuperscript{m}45\textsuperscript{s}.3+76$ ^{\circ} $53\arcmin 09\arcsec}} & 
		\multicolumn{1}{p{0.29in}}{\centering 15.7} & 
		\multicolumn{1}{p{0.38in}}{\centering 0.131} & 
		\multicolumn{1}{p{0.48in}}{\centering H$ \beta $ } & 
		\multicolumn{1}{p{0.7in}}{\centering SED550 \par \centering SED525} \\
		\hline
		\hline
		
	\end{tabular}
\label{tab:obj}
\end{table*}

One of the most reliable methods for measuring the masses of the central black holes in AGNs is reverberation mapping (\citealt{BlanfordMKee82}) of their central regions. The mass is determined by assuming the emitting gas in the broad line region (BLR) to be virialized:
\begin{equation}
\label{formula:virial}
M_{\rm SMBH}=f \times (R_{\rm BLR} \vartheta_{\rm line}^2G^{-1}),
\end{equation}
where $ G $ is the gravitational constant,
$R_{\rm BLR}$ is the BLR size,
$ \vartheta_{\rm line} $ is the velocity of the line-emitting gas in the BLR, and $ f $ is a dimensionless factor of the order of unity dependent on the BLR structure and kinematics as well as the system’s inclination relative to the observer \citep[see.,~e.g.,][]{Peterson04}. 
The BLR size is defined as $ R_{\rm BLR} \equiv c \tau $, where $ c $ is the speed of light and $\tau$ is the time lag of the radiation in the emission line forming in the BLR relative to the continuum radiation from the accretion disk \citep{Peterson93}. The H$\beta$ 4861 \AA{} line is most commonly used for reverberation mapping \citep{Bentz13}, although, for example, in the pioneering paper by \cite{Cherep73} the study was carried out in the H$\alpha$ 6563 \AA{} line. For more distant galaxies the radiation time lag is investigated in the Mg\,II 2798 \AA{} \citep{Homayouni20,Za} and C\,IV 1549 \AA{} \citep{ShenCIV} lines.

For a reliable determination of the time lag between the line and continuum light curves, it is necessary to conduct a long-term AGN monitoring (of the order of several years for typical sizes $R_{\rm BLR}$ $\sim$  0.02 pc), which is not a simple task in terms of the expenditure of telescope time. The photometric reverberation mapping method in medium- and narrow-band filters has gained in popularity in the last decade against the background of the spectroscopic method \citep{Haas11}. 

Since 2018 a photometric monitoring of an AGN sample predominantly at the 1-m Zeiss-1000 telescope of SAO RAS \citep{Zeiss} has been conducted within our reverberation mapping program. In this paper, we present the results of observations for the two brightest galaxies with broad lines from the sample under study \citep{RM_1}, 2MASX J08535955+7700543 (hereafter J0853+77) and VII Zw 244: their light curves and detected variability time lags. Spectroscopic data were obtained at BTA to estimate the gas velocities. The result of this work is the measurement of the central SMBH masses in the nuclei of the two studied galaxies, with $M_{\rm SMBH}$ having been estimated previously for these objects based only on indirect methods.

\section{OBSERVATIONS}

The characteristics of the studied galaxies are given in Table \ref{tab:obj}. Our observations were carried out with the SAO RAS telescopes: 1-m Zeiss-1000 and 6-m BTA. The first one was used for monthly photometric observations; at BTA spectroscopy was performed and several photometric observational epochs were acquired.

\subsection{Photometry}

\begin{figure*} 
	\centering
	\includegraphics[width=1\linewidth]{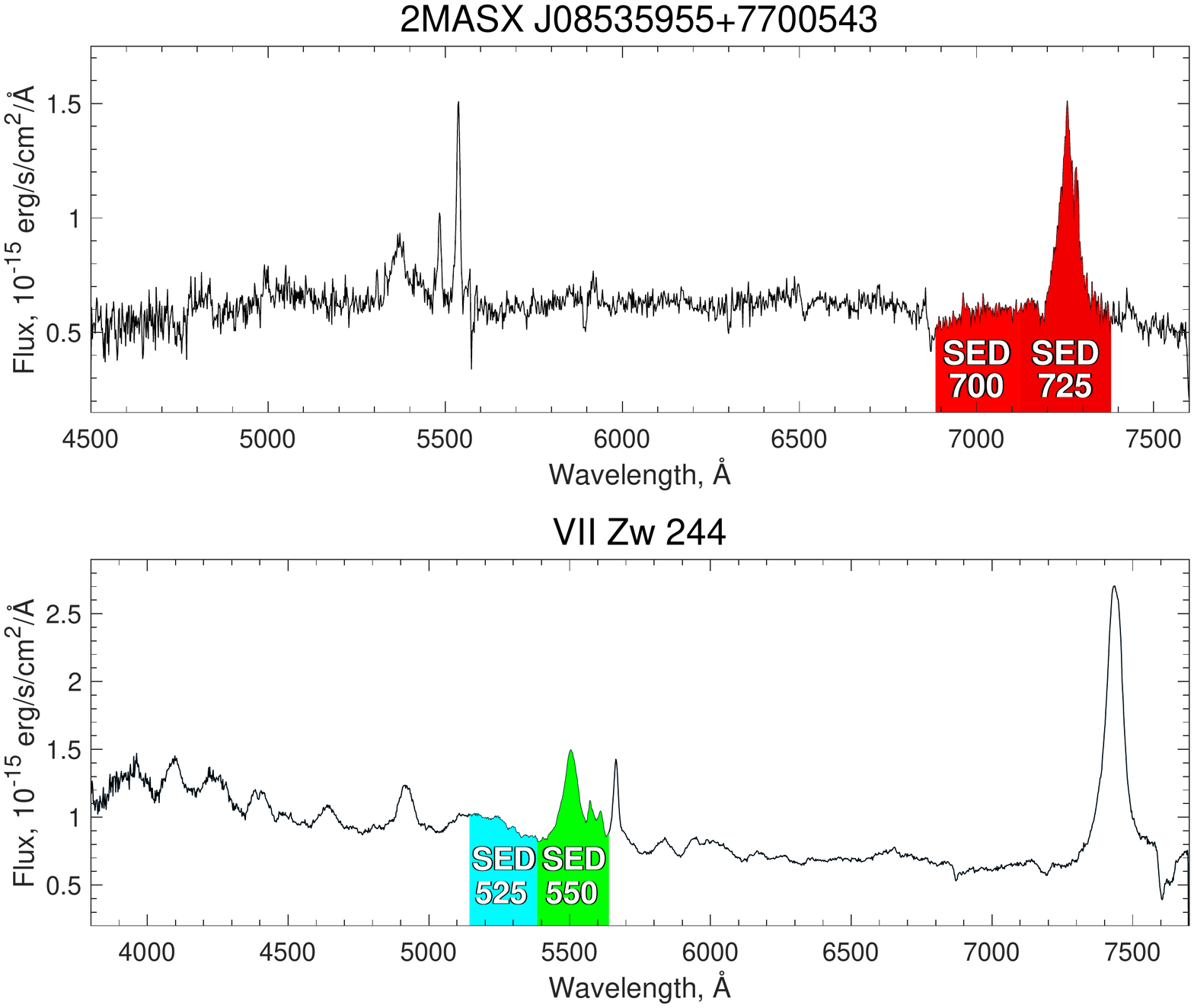}
	\caption{BTA spectra of J0853+77 and VII Zw 244. The pass bands of the filters used are  overplotted.}
	\label{spec}
\end{figure*}

The observations of each object imply the use of two interference filters: one corresponds to the region of the broad Balmer emission line and the other corresponds to the continuum in the region close to the line. Medium-band filters with typical bandwidths $\sim$250 \AA\ are used in the experiment. The spectral range of the galaxies being studied covered by the filters used is illustrated in Fig.~\ref{spec}.

Three instruments mounted at the Cassegrain focus were successively used during the monitoring at the 1-m telescope:
\begin{itemize}
    \item MaNGaL \citep{MaNGaL20} + Andor iKon-M 934/Andor Neo sCMOS (2560$\times$2160);
    \item MMPP \citep{MMPP}
    + Eagle V (2048$\times$2048);
    \item StoP \citep{stop} + Andor iKon-L 936 (2048$\times$2048).
\end{itemize}

To obtain additional photometric data, we performed observations at BTA with the SCORPIO-2 focal reducer \citep{SCO2} + E2V 42-90/E2V 261-84. 
\begin{figure}[!th]
	\begin{center}
		\includegraphics[width=8cm]{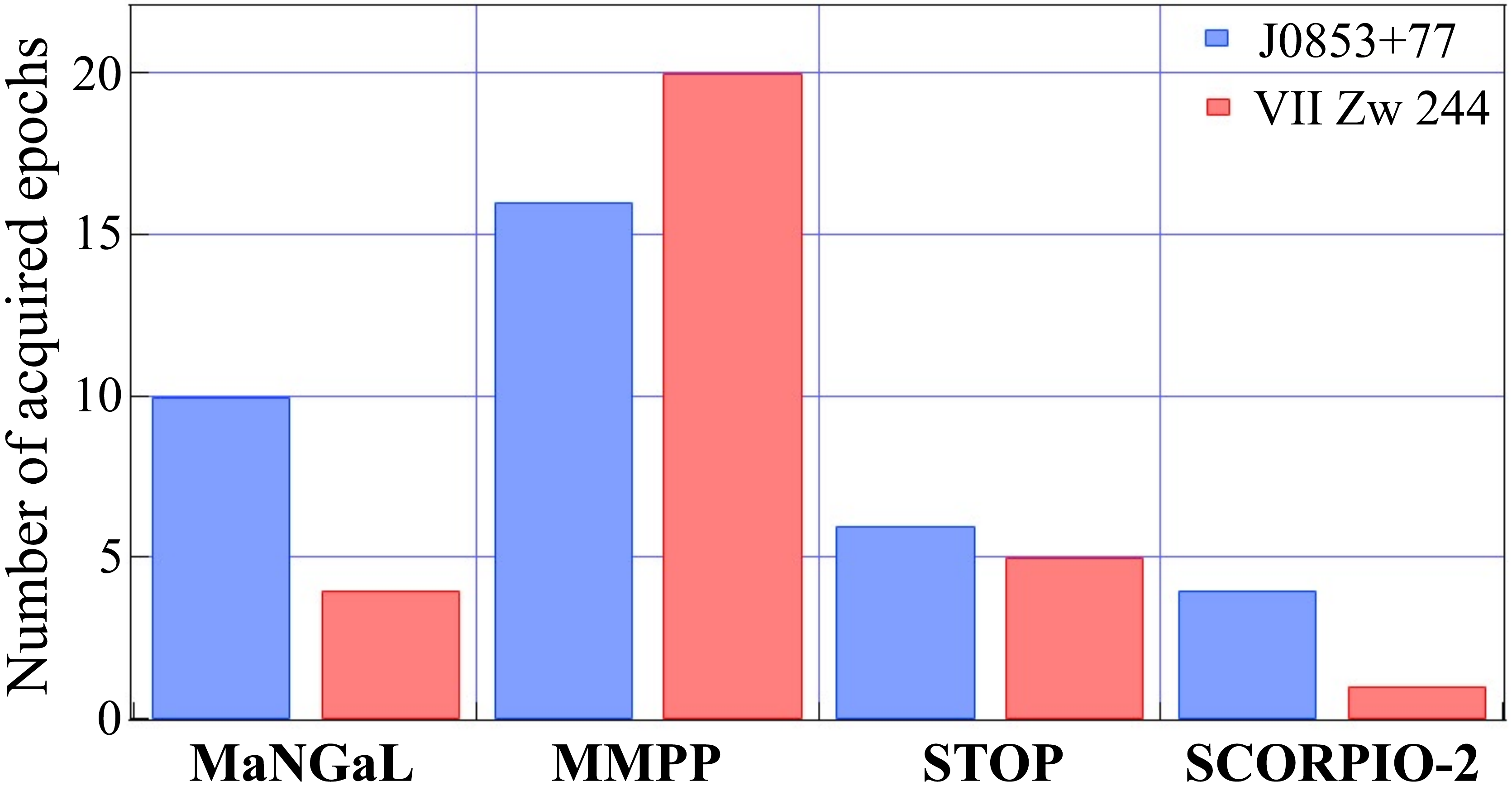}
		\caption{The amount of photometric data obtained with various instruments.}
		\label{fig:devices}
	\end{center}
\end{figure}

Figure~\ref{fig:devices} presents a histogram illustrating the number of observed epochs for each studied object on various instruments. A total of 36 and 30 observing epochs were acquired for the galaxies J0853+77 and VII Zw 244, respectively. The monitoring durations for J0853+77 and VII Zw 244 are 814 and 610 days, respectively.

\begin{figure*}[h]
	\begin{center}
	    \includegraphics[width=17.3cm]{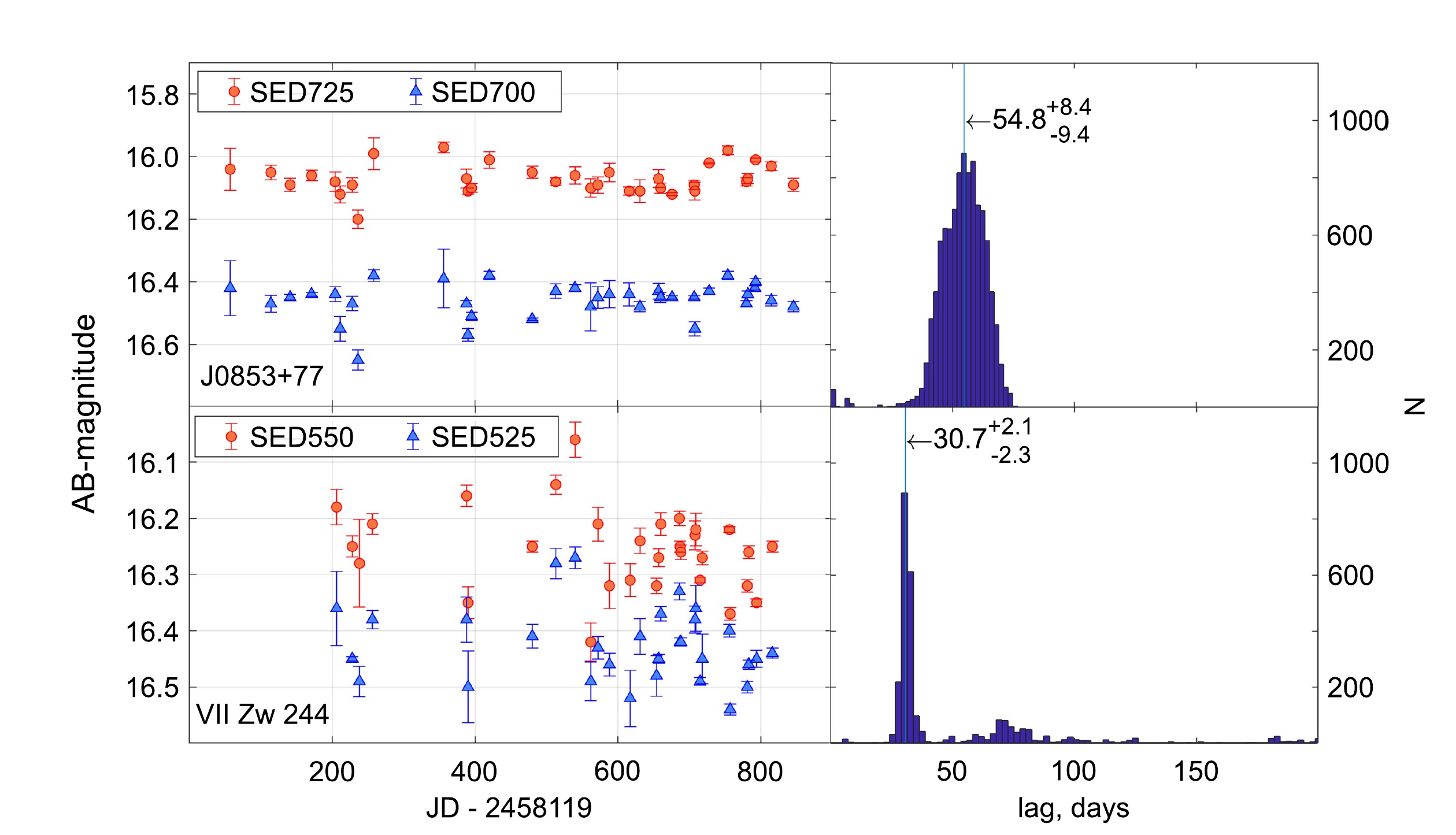}
		\caption{Light curves (left) and the corresponding distributions of lags $ \tau$  (right) for J0853+77 and VII Zw 244. The blue triangles represent the photometric AB magnitude measurements in the continuum near the line; the red circles correspond to the photometric AB magnitude measurements in the spectral region with a broad Balmer emission line. The Julian dates are counted from January 1, 2018.}
	\label{fig:LCjav}	
	\end{center}
\end{figure*}

The technique of photometric observations is described in detail in \citet{serb}. The constructed AGN light curves are shown in Fig.~\ref{fig:LCjav} (left).

\subsection{Spectroscopy at BTA}

\subsubsection{VII Zw 244}

A low-resolution spectrum of VII Zw 244 was taken on October 9, 2019, at the 6-m BTA telescope using SCORPIO-1 \citep{SCO1} in the long-slit mode to update the previous spectroscopic data obtained more  than 25 years ago by \cite{BorosonGreen92}. Moreover, the spectrum from \cite{BorosonGreen92} spans the range 4070–-5883~\AA \AA; in this paper the spectrum was extended to the range 3700--7700~\AA \AA. 

The spectrum was taken with a 1"{}-wide slit and VPHG\,550G (volume phase holographic grating). The reciprocal dispersion is 2.1~\AA/pixel and the instrumental profile is FWHM 10~\AA. The BTA prime focus adapter was used to calibrate the spectrum \citep {adap}. The comparison spectrum from a He-Ne-Ar lamp and flat fields were taken at the same telescope position. Three 5-min exposures were taken. The spectrophotometric standard G191B2B was observed  on the same night at a similar zenith distances to correct the sensitivity curve of the~E2V42-40 CCD and to minimize the spectral atmospheric transmission effects. To take into account the light losses on the slit and then to estimate the total flux, we took a slitless spectrum of the object.

The resulting spectrum of VII Zw 244 is presented in Fig.~\ref{spec} (bottom).

\subsubsection{2MASX J08535955+7700543}

We took a spectrum of the AGN J0853+77 on November 6, 2019, at BTA using the SCORPIO-2 focal reducer with the E2V42-90 CCD in the long-slit mode to update the spectroscopic data obtained by \cite{Wei99} more than 20 years ago.

The spectrum was taken with a 2"{}-wide slit and VPHG\,940@600 (reciprocal dispersion 1.16~\AA/pixel) {in the spectral range 4200--7700~\AA \AA.} Four 15-min exposures were taken. The comparison spectrum and flat fields were also taken with the prime focus adapter. We corrected the CCD sensitivity curve based on the observations of the standard G191B2B. The spectrum is presented in Fig.~\ref{spec} (top). 

The accuracy of the spectroscopic data turned out to be insufficient when analyzing the H$\alpha$ profile (see below). Therefore, in addition to the spectroscopic data, we used  the spectropolarimetric data obtained on March 3, 2020, with the SCORPIO-2 focal reducer. In the spectropolarimetric mode VPHG1026@735 and a double Wollaston prism were inserted into the beam and spectra were simultaneously recorded on the detector in four polarization directions: 0$^{\circ}$, 90$^{\circ}$ and 45$^{\circ}$, 135$^{\circ}$. A blocking GS-17 filter was used to suppress the second grating order. The slit width was 2". At the same telescope position we took the frames of flat fields, a comparison spectrum, and a 3-point test to correct the field geometry. Our observations were carried out in a series of 10 exposures (300 s + 9 $\times$ 600 s). The long total exposure allowed a high signal-to-noise ratio both in the lines and in the continuum to be achieved for the observed object after the addition of all polarization directions. These spectroscopic data were used to decompose the profile and are presented in Fig. \ref{fig:lineDecomposition} (right).

\section{DATA REDUCTION AND ANALYSIS}
\label{sect:analisys}

\subsection{Photometric Data}

For a better photometric accuracy needed for AGN variability studies, we used the method of differential photometry relative to the local standards in the object’s field of view. Independent processing of each frame \cite[for details, see][]{RM_1} provides a typical photometric measurement error $0^{\rm m}.01$--$0^{\rm m}.03$. The secondary standards are given in \cite{RM_1}. The continuum and line light curves for VII Zw 244 and J0853+77 are presented in Fig.~\ref{fig:LCjav} (left).


\begin{figure*}[h]
	\begin{center}
	    \includegraphics[width=16.3cm]{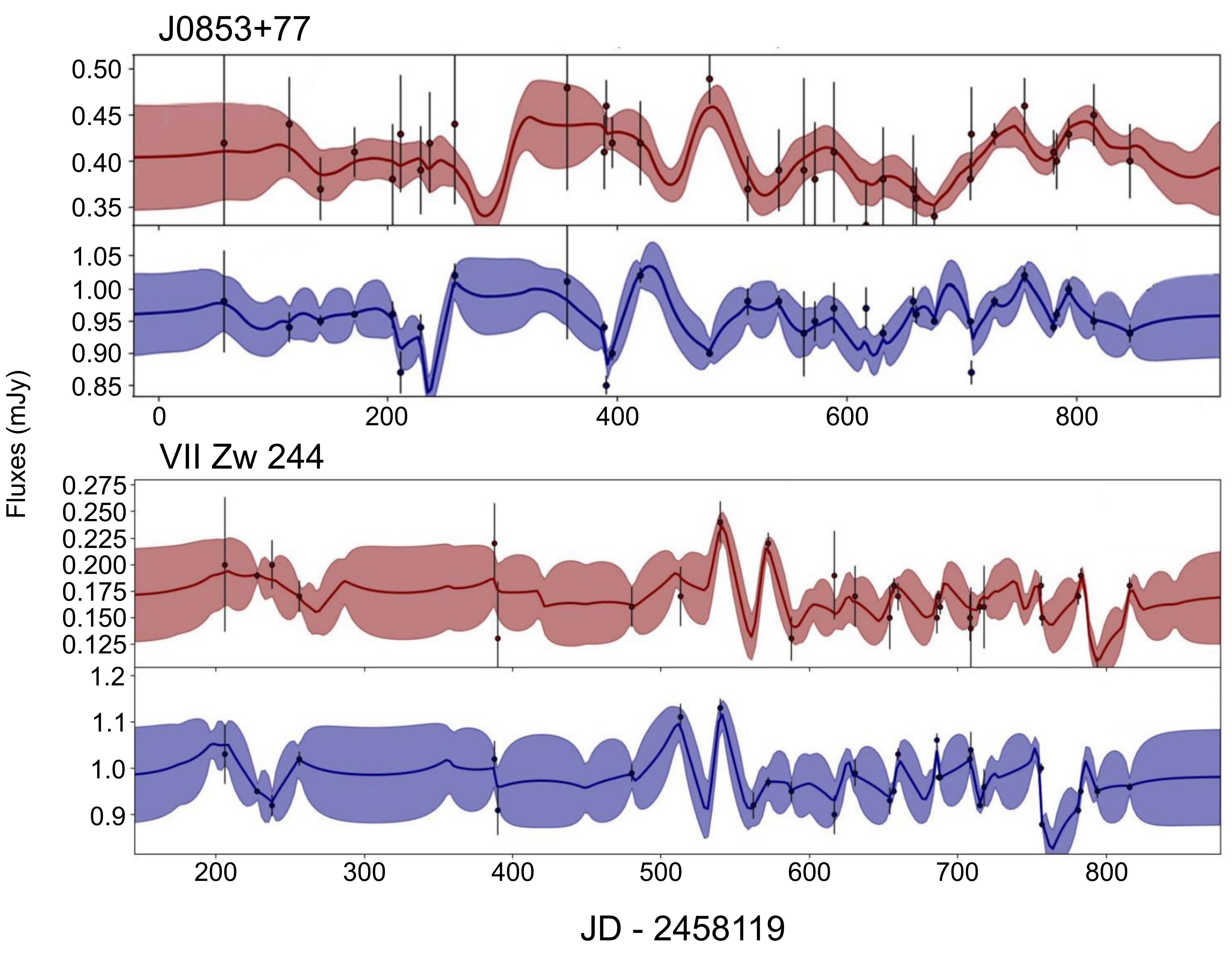}
		\caption{JAVELIN simulations of the light curves for J0853+77 and VII Zw 244. The black dots mark the original observational data with error bars. The red and blue curves describe the most probable model of emission line radiation variations with the subtracted continuum and the most probable model of continuum radiation variations, respectively. The pale-red and pale-blue regions correspond to the regions of admissible values for the model light curves. The Julian dates are counted from January 1, 2018.}
	\label{fig:modeljav}	
	\end{center}
\end{figure*}

To estimate the time lag between two light curves for each object, we applied the method using the JAVELIN code 
\citep{Zu16,Yu20} implemented in the Python programming language. The results in the form of histograms are presented in Fig.~\ref{fig:LCjav} (right), where $N$ is the frequency of occurrence of the sets of parameters during the MCMC sampling. The simulations used 10 000 sets of parameters. The simulated light curves are presented in Fig.~\ref{fig:modeljav}. The derived lags are $ \tau = 54.8_{-9.4}^{+8.4}$  days for J0853+77 and $ \tau  = 30.7_{-2.3}^{+2.1}$  days for VII Zw 244.

\subsection{Spectroscopic Data}

The spectroscopic data for VII Zw 244 and J0853+77 were reduced using the package implemented in the IDL\footnote {Interactive Data Language, \url{https://www.harrisgeospatial.com/Software-Technology/IDL} } environment and provided by S.N.\,Dodonov. The data reduction implies the following: bias subtraction, cosmic-ray particle hit removal, flat fielding, wavelength calibration, night-sky line subtraction, correction for the atmospheric and spectrograph transmission  based on spectrophotometric standards, and extraction into a one-dimensional spectrum. 

For the most proper absolute flux calibration, the spectra were convolved with the transmission curves of the filters used in our photometric observations. The derived synthetic fluxes in the filters were calibrated to real photometric observations with a higher accuracy than that of the spectroscopic ones from the standpoint of absolute values. Thus, we achieved
complete agreement between the absolute fluxes measured in both photometric and spectroscopic observations.

To analyze the H$\alpha$ profile in the spectrum of J0853+77, we used the integrated spectropolarimetric data, which give a higher signal-to-noise ratio. The data reduction includes a standard procedure for long-slit spectroscopy: bias subtraction, flat fielding, geometrical correction along and across the slit,
night-sky subtraction, correction for the spectral sensitivity of the instrument, and spectral wavelength calibration. The method of observations and the data reduction are described in detail in \cite{Afanasiev12}.

After the data reduction all four polarization directions were added to obtain the integral energy distribution in wavelengths. However, the dispersion curve correction and the absolute spectrum calibration turned out to be unsatisfactory and, for this reason, the complete spectrum is not provided.  This spectrum is used to analyze the H$\alpha$ profile (Fig.~\ref{fig:lineDecomposition}). The absolute fluxes were reconstructed in the same way as in the case of spectroscopic data.

\section{SMBH MASS ESTIMATION}

\begin{figure*}[h]
	\begin{center}
	    \includegraphics[width=0.9\linewidth, angle=180]{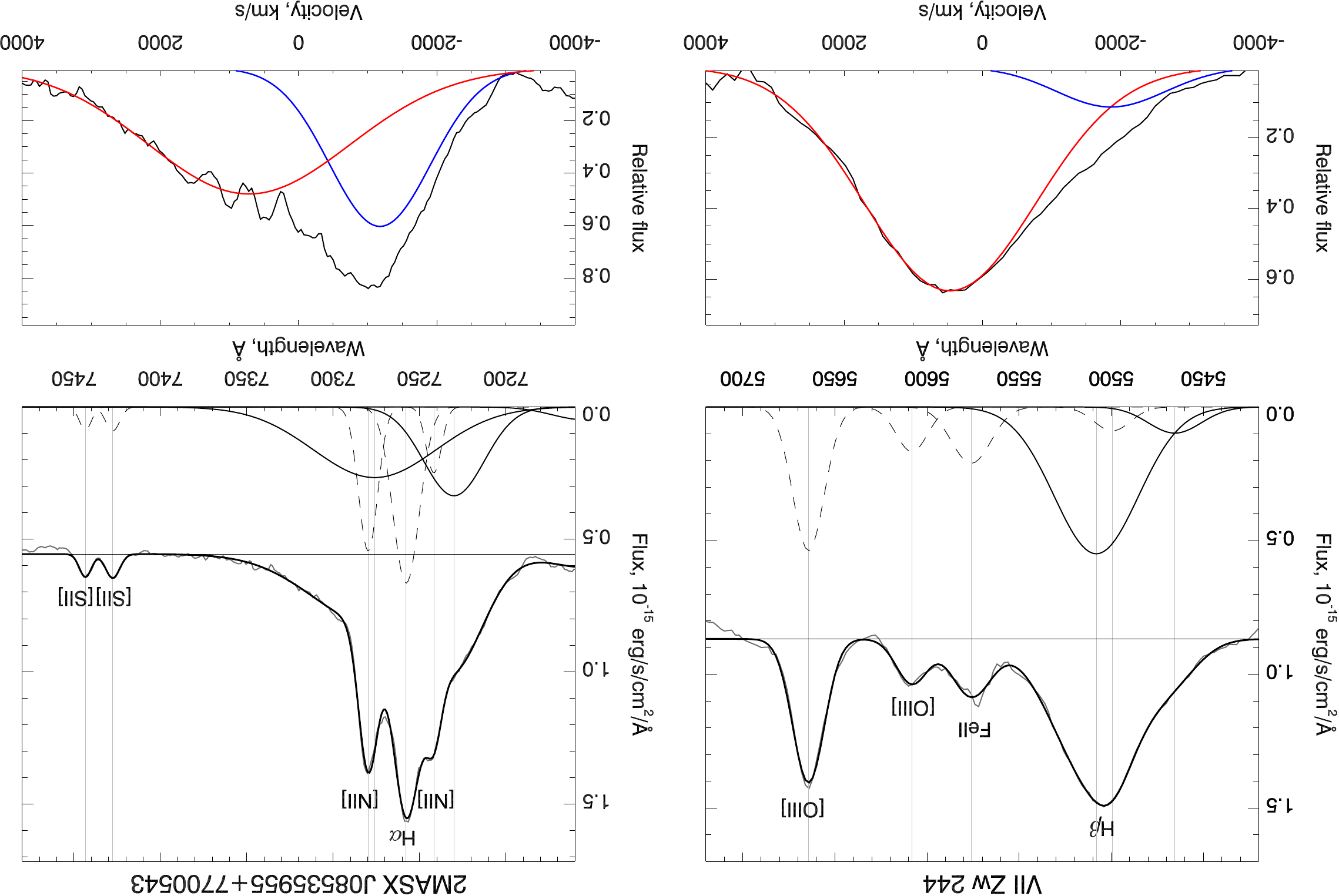}
		\caption{Decomposition of the observed emission line profiles. Upper panels: the H$\beta$ region in the spectrum of the galaxy VII Zw 244 and its decomposition into Gaussian components (left) and the H$\alpha$ line region in the spectrum of the galaxy J0853+77 and its decomposition into Gaussian components (right). The narrow components of the [O III], Fe II and H$\beta$ lines for VII Zw 244 and the [N II] and H$\alpha$ lines for J0853+77 are indicated by the dashed lines. Lower panels: the profiles of the H$\beta$ line in the spectrum of the galaxy VII Zw 244 (left) and the H$\alpha$ line in the spectrum of the galaxy J0853+77 (right) with the subtracted narrow  components. The red and blue lines indicate the broad Gaussian profiles shifted toward positive and negative velocities, respectively.}
		\label{fig:lineDecomposition}
	\end{center}
\end{figure*}

To estimate the masses, it is necessary to estimate the gas velocities in the BLR. For this purpose, the first step in analyzing the emission line profile was to subtract all narrow components and to obtain a pure broad component of H$\alpha$ for J0853+77 and H$\beta$ for VII Zw 244. The profile was analyzed by the multi-Gaussian decomposition method. 

The decomposed H$\alpha$ profile for J0853+77 is shown in Fig.~\ref{fig:lineDecomposition} (upper right panel). As can be seen from the figure, the profile of the observed emission line consists of two narrow [N II] doublet lines, a
narrow H$\alpha$ component, and a broad H$\alpha$ component fitted by two Gaussians. The final profile of the broad component minus the narrow lines is presented in Fig.~\ref{fig:lineDecomposition} (lower right panel). Let us calculate the second moment of the velocity  $<\vartheta^{2}>$ from the broad H$\alpha$ profile:
\begin{equation}
\label{formula:vel}
< \vartheta ^{2}> =\frac{ \sum _{}^{} \left(  \vartheta -< \vartheta > \right) ^{2} \times f \left(  \vartheta  \right) }{ \sum _{}^{}f \left(  \vartheta  \right) },  
\end{equation}
where  $f(\vartheta)$ is the line flux in the scale of velocities $\vartheta$ and the first moment of the velocity $<\vartheta>$ is defined as
\begin{equation}
< \vartheta> =\frac{ \sum _{}^{} \vartheta \times f \left(  \vartheta  \right) }{ \sum _{}^{}f \left(  \vartheta  \right) }. 
\end{equation}
Then, the gas rotation velocity in the BLR in the observed H$\alpha$ line for J0853+77 is then taken to be  
$$\vartheta^{2}_{\rm H\alpha} =\  <\vartheta^{2}>\  = 2.3 \times 10^{6} ~ \textrm{km}^{2}/\textrm{s}^{2}.$$

A similar analysis using a multi-Gaussian decomposition was also made for the observed H$\beta$ line in the object VII Zw 244. The profile decomposition is presented in Fig.~\ref{fig:lineDecomposition} (upper left panel). The contribution of the narrow [O III], Fe II lines and narrow H$\beta$ component were subtracted from the H$\beta$ profile. The resulting broad profile fitted by two Gaussians is presented in Fig.~\ref{fig:lineDecomposition} (lower left panel). The second moment of the velocity was calculated from Eq. (\ref{formula:vel}). The gas velocity in the H$\beta$ line for VII Zw 244 was
estimated to be
$$\vartheta^{2}_{\rm H\beta} =\  <\vartheta^{2}>\  = 1.9 \times 10^{6} ~ \textrm{km}^{2}/\textrm{s}^{2}.$$ 

Now, having determined the gas velocity in the region emitting the observed emission line and the distance to the emitting region measured by the reverberation mapping method, let us estimate the SMBH mass for the two galaxies from Eq. (\ref{formula:virial}). The dimensionless factor $f$ will be assumed to be equal to unity. For the galaxy J0853+77 $M_{\rm SMBH} \sim2.5\times 10^{7} \  M_{\odot}$. For the galaxy VII Zw 244 the measurement showed $M_{\rm SMBH} \sim1.1\times 10^{7} \  M_{\odot}$. 


\begin{table*}[h]
	\centering
	\caption{Estimates of the parameters for the central AGN regions. From left to right, the table gives: the object name, the response lag in the emission line $\tau$ in days, the square of the gas velocity $\vartheta^2_{\rm line}$ measured from the emission line profile, the object luminosity at 5100 \AA\ $\lambda L_{5100}$, our estimate of the SMBH mass $M_{\rm SMBH}$, and the estimate of the SMBH mass given in $^*$\cite{Xu07} and
        $^{**}$\cite{Tilton13} }
	\begin{tabular}{lccccc}
		\hline
		\hline
	Object & $\tau$,  & $\vartheta^2_{\rm line}$,  &  $\lambda L_{5100}$  & $M_{\rm SMBH}$ & $M_{\rm SMBH}^{\rm calib}$ \\
	& days & $10^6$ km$^2$/s$^2$ & $10^{44}$ erg/s & lg($M/M_{\odot}$) & lg($M/M_{\odot}$) \\
    \hline
    J0853+77 & $54.8_{-9.4}^{+8.4}$ & 2.3 & 0.87  & $7.398^{+0.153}_{-0.171}$  & 8.228$^*$ \\
    \hline
    VII Zw 244 & $30.7_{-2.3}^{+2.1}$ & 1.9 & 1.67 & $7.049^{+0.068}_{-0.075}$  & $7.825^{+0.087}_{-0.109}$ $^{**}$ \\
            \hline
            \hline
        
	\label{tab:Mbh}
	\end{tabular}
\end{table*}


Before this study the SMBH masses for these objects have been estimated only by indirect methods. The SMBH mass in the object J0853+77 was estimated by \cite{Xu07} based on the spectrum from \cite{Wei99} using the BLR size–-luminosity calibration relation from \cite{Kaspi00}. The estimate was $M_{\rm SMBH} \sim1.7\times 10^{8} \  M_{\odot}$, differing from the one in this paper by a factor of $\sim$7. This difference can be explained by two factors. The first of them is an insufficient quality of the spectrum used in \cite{Xu07}, which did not allow a detailed analysis of the line profiles to be performed and the gas velocity to be determined with a minimal error. The uncertainty introduced by the luminosity determination from the spectroscopic data, which we avoided by tying the spectroscopic data to the photometric ones, can also be significant. 

\cite{Tilton13} also estimated the SMBH mass in the galaxy VII Zw 244 indirectly using the BLR size–-luminosity calibration from \cite{Bentz09} and the spectroscopic data from \cite{BorosonGreen92}. The estimate given in the paper was $M_{\rm SMBH} = (6.7\pm1.5) \times 10^{7} \  M_{\odot}$. Although this estimate differs from our one by a factor of $\sim$6, it is close within the error limits. Probably, the spectra in \cite{BorosonGreen92} did not allow a detailed analysis of the H$ \beta$ profile to be performed either. 

The derived characteristics of the objects investigated in this paper, J0853+77 and VII Zw 244, are summarized in Table~\ref{tab:Mbh}. The table presents the response lags in the emission lines $\tau$ equal to the BLR size in light days, the gas velocity $\vartheta^2_{\rm line}$ determined by analyzing the line profiles, and the masses in units of $\lg (M/M_{\odot})$ estimated in this paper, $M_{\rm SMBH}$, and in the earlier papers of other authors, $M_{\rm SMBH}^{\rm calib}$ .

An empirical BLR size–-luminosity relation is one of the goals of the reverberation mapping program. The AGN luminosity in this case is determined either in an emission line or at some effective wavelength in a continuum free from emission lines. Table~\ref{tab:Mbh} gives the luminosities of the two studied galaxies at 5100 \AA. The flux $f_{\lambda}$ was integrated in the range 5000--5200 \AA\AA{} in the galaxy’s frame of reference and was divided by the width of the integration window to obtain the monochromatic luminosity. The residual influence of the H$\beta$, Fe II, and approximately constant [O III] profiles was first subtracted from the spectrum in this region. The luminosity is then defined as
$$
\lambda L_{5100} = 4 \pi D^2 f_{\lambda}, 
$$
where $D$ is the AGN distance. The Hubble constant was taken to be $H_{0}=67.4$~(km/s)/Mpc. The contribution of the host galaxy was not subtracted. The luminosities of the galaxies are $\lambda L_{5100} = 0.87 \times 10^{44}$ erg/s for J0853+77 and $\lambda L_{5100} = 1.67 \times 10^{44}$ erg/s for VII Zw 244. Based on the luminosity estimates at 5100 \AA\, we estimated the bolometric luminosities of the galaxies. The bolometric correction $BC$ was taken from \cite{rich}: $BC = 9.26$. The luminosity estimates for the galaxies are $L_{\rm bol} \approx 8.1 \times 10^{44}$ erg/s for J0853+77 and $L_{\rm bol} \approx 15.5 \times 10^{44}$ erg/s for VII Zw 244, which are 0.27$L_{\rm Edd}$ and 1.17$L_{\rm Edd}$ ($L_{\rm Edd}$ is the Eddington luminosity) for the galaxies, respectively. It is worth noting that this fraction of the bolometric luminosity of the Eddington one, first, is overestimated a priori due to the nonsubtracted contribution of the host galaxy and, second, depends on the chosen $BC$. For instance, using $BC = 7.79$ from \cite{kraw}, the luminosities are estimated to be 0.22$L_{\rm Edd}$ for J0853+77 and 0.98$L_{\rm Edd}$ for VII Zw 244, which is particularly critical for the second galaxy. On the whole, however, the bolometric luminosities for the investigated objects turn out to be high and close to the critical ones, which arouses interest in a more careful study of the AGNs J0853+77 and VII Zw 244.

\section{CONCLUSIONS}
\label{sect:discussion}

Our monitoring of the galaxies J0853+77 and VII Zw 244 for more than two years at the SAO RAS telescopes using a JAVELIN analysis allowed the BLR sizes $R_{\rm BLR}$=c$ \tau$ in these AGNs to be estimated: $54.8_{-9.4}^{+8.4}$ and $30.7_{-2.3}^{+2.1}$ light days or $0.046_{-0.008}^{+0.007}$  and  $0.026_{-0.002}^{+0.002}$  parsecs, respectively. We estimated the gas velocities in the BLRs from the spectroscopic data obtained at the 6-m SAO RAS BTA telescope. Based on new observational data, we  estimated the SMBH masses at the centers of the galaxies 2MASX J08535955+7700543 and VII Zw 244 to be 
		$\lg(M/M_{\odot}) = 7.398_{-0.171}^{+0.153}$   and 
		$\lg(M/M_{\odot}) = 7.049_{-0.075}^{+0.068}$,  respectively.

\paragraph{ACKNOWLEDGMENTS}  
This work was supported by RSF grant no. 20-12-00030 "Investigation of the Ionized Gas Geometry and Kinematics in Active Galactic Nuclei by Polarimetry Methods".
We also thank V.L. Afanasiev, A.N. Burenkov, and S.N. Dodonov for useful discussions and remarks. The observations at the SAO RAS telescopes are  performed with support from the Ministry of Science and Higher Education of the Russian Federation (including contract no. 05.619.21.0016, unique project identifier RFMEFI61919X0016).



\end{document}